\documentstyle[sprocl]{article}

\begin{document}

\title{ PROBABILITY AND THE MANY WORLDS INTERPRETATION OF QUANTUM THEORY}

\author{L. VAIDMAN}

\address{Centre for Quantum Computation,\\
 Department of Physics, University of Oxford,\\
 Clarendon Laboratory, Parks Road, Oxford OX1 3PU, England\\and\\
 School of Physics and Astronomy,\\
Raymond and Beverly Sackler Faculty of Exact Sciences\\ 
Tel-Aviv University, Tel-Aviv 69978, Israel}

\maketitle
\abstracts{Two works related to the concept of probability
  in the framework of the many-worlds interpretation are
  presented. The first deals with recent controversy in classical
  probability theory. Elga and D. Lewis argues that Sleeping Beauty
  should have different  credences for result of a fair coin toss in a
  particular situation. It is argued that when the coin is replaced by
  a quantum coin, the credence is unambiguous and since one should not
  expect a difference between classical and quantum coins, this
  provides a particular resolution of the controversy. Second work
  is an analysis of a recent criticism by Byrne and Hall of
  Everett-type approach presented by Chalmers. It is shown that the
  criticism has no universal validity and that the Byrne and
  Hall rejection of any Everett-type interpretation of quantum mechanics is unfounded.}

\section{Introduction}
The talk I presented in this conference analyzed the meaning of recent
experiments claiming to close last loopholes in testing Bell's
inequality. I argued that the really important task in this field has
not been tackled yet and that the leading experiments claiming to
close locality and detection efficiency loopholes, although making
very significant progress, have conceptual drawbacks. The important
task is constructing quantum devices which will allow winning games of
certain correlated replies against any classical team, and I proposed
a novel game of this type named an ``impossible necklace game''.
However, this work has been recently published (Vaidman, 2001a), so I
present here, instead, two recent unpublished works which I discussed
extensively in the conference (mostly walking around the beautiful
Vaxjo lake).  

The two works presented here are closely related to the main theme of the
conference: the meaning of probability in quantum mechanics. The first
one was done together with Simon Saunders (see the discussion of the
classical case in Vaidman and Saunders 2001) and it is presented in
Section 2. It analyzes recent controversy in classical probability
theory by transforming it to the framework of quantum theory. It is
surprising that quantum theory, with its serious difficulties in
dealing with the concept of probability, can help in the analysis of a
classical probability problem. The second work (presented in Section
3) defends the many-worlds approach to quantum mechanics against
recent criticism by Byrne and Hall of the Chalmers book that presented
the many-worlds interpretation of quantum mechanics as the most
promising approach.

Before going to these sections I want to express my understanding of
the main discussion in this conference. A very significant part of the
conference was allocated to the analysis of Bayesian probability in the
framework of quantum theory (see, e.g., Caves, Fuchs, and Schack,
2001). In my rough understanding, Bayesian approach denies existing
underlying ontology and defines probability from several axioms
related to properties of the concept of probability. In the conference
we witnessed a very impressive progress in the Bayesian probability
program. However, many participants were not ready to accept this
approach due to difficulties of physics without ontology. I am fully
accept the criticism of frequency approach to probability, but I am
not ready to give up the ontology.

 My ontology is the many-worlds
ontolgy (Everett 1957). It turns out that in spite of having the
ontolgy, I do need some kind of Bayesian approach to probability. The
many-worlds interpretation has a serious difficulty with dealing with
probability and my solution of this difficulty (Vaidman 1998, 2001b)
uses an approach that adopts most if not all principles of Bayesian
approach. So if before the conference I thought that Bayesian approach
is orthogonal to my views on probability, now I consider it as a
complementary one. The many-worlds approach resolves the difficulties
of satisfactory description of the collapse of a quantum wave and the
Bayesian approach solves the difficulty of the concept of probability
in a deterministic theory such as the many worlds interpretation in
which all possible outcomes of experiments are actual.

\section{Sleeping Beauty in Quantumland}

\subsection{Lewis - Elga  controversy}

 Lewis's comment (2001) on Elga's paper (2000) has far reaching
 consequences. If Lewis is right, then the approach to credence for an
 event as the value of an ``intelligent bet'' on this event (e.g.
 Sklar 1993) does not have universal applicability.  The betting
 approach to this question (Aumann et al.  1997) leads to Elga's
 result which Lewis contests.  I believe, however, that Lewis's
 approach is untenable, and thus the universality of the betting
 approach to probability has not been breached.

I give two arguments in favour of Elga's conclusion, the first classical, the
second quantum mechanical. It is of interest that a quantum mechanical
analysis can be brought to bear on this classical probability problem.

The first is as follows: consider the following three experiments, the first
of which is Elga's. I claim the credence of Beauty is the same
in each case.

\begin{quotation}
(i) Beauty sleeps during the week. On Sunday a fair coin is tossed. If the
result is Heads (H), Beauty is awaken on Monday only, if the result is Tails
(T), Beauty is awaken on Monday and Tuesday. On every awakening, Beauty is
told nothing, but must answer the question: ``What is your credence for the
coin to be H?'' After the conversation her memory of the awakening is erased
and she sleeps again.

(ii) Beauty sleeps for 100 years, uninterrupted save for awakenings according
to procedure (i), which take place every week, following a coin toss on each
Sunday of each week.

(iii) Beauty sleeps for 100 years, uninterrupted save for 7827 awakenings.
Using a classical random number generator to determine the order, on 2609
awakenings a coin is placed H, and on 5218 awakenings a coin is placed T. On
each awakening, the same procedure is followed as in (i).
\end{quotation}

Experiment (ii) is not just a repetition of experiment (i) 5218 times.
In the latter, on every awakening, Beauty knows which week it is. In
(ii) Beauty does not know which week it is and, therefore, she does
not know which coin in the sequence the question is about. However,
since all the coins are fair, Beauty's situation is the same in all
weeks and thus the lack of information which week it is cannot make a
difference in her credence.

Experiments (ii) and (iii) are also not identical. The probability in
experiment (ii) that one will obtain exactly 2609 H will is very
small.  However, the probability that this number will be different
from 2609 by more than, say, 100, is also very small. Therefore, the
probability that the relative frequency of H will be significantly
different from 1:2 is negligible.  Thus, although I do not have an
argument according to which Beauty has to give exactly the same answer
in (ii) and (iii), I can argue that these answers cannot differ
significantly.  Since our job is to decide between credences 1/2
(Lewis) and 1/3 (Elga), this is enough.

In experiment (iii) it is obvious that Beauty should give credence 1/3 for H.
If there is no significant difference in Beauty's answer in case (iii) and
(ii), and no significant difference in her answer in case (ii) and (i), then
in Elga's experiment Beauty should give the answer 1/3, and not 1/2 as Lewis claims.

There is a conceptual difference between (i) and (ii) (and a similar
difference between (i) and (iii)). In (i) the Beauty was asked a question
about an uncentered proposition: ``What is the state of the coin?'' In
contrast, in (ii) Beauty was asked a question about a centered proposition:
``What is the state of the coin of {\it this} week?'' The unusual feature
of Elga's experiment, that Beauty must alter her credence about an uncentered
proposition with no new uncentered evidence, is not present here. But  this
does not alter the fact that Beauty must give the same answer in case (ii) and
(i).

Apart from the small statistical difference between (ii) and (iii) already
mentioned, the two differ in another aspect. In (ii) there is something which
corresponds to Lewis's number 1/2: in half of the weeks of Beauty's sleep the
coin is H, and \ in half it is T. In contrast, in experiment (iii), nothing
corresponds to 1/2. However, knowledge of this statistical structure to her
string of awakenings, in case (ii), is not knowledge that Beauty can use,
since never on awakening does she learn where in the string she is located.

\subsection{The inconsistency of the  Lewis approach}

In order to see how the difference between the one-week experiment and the
many-weeks experiment arises, and in order to show the inconsistency of
Lewis's approach,
 consider an
experiment of the kind (ii) but limited to two weeks. A fair coin is tossed
twice. The credence of Beauty $p(H)$ can be calculated as the sum of
conditional probabilities on the outcomes of the coin tosses:
\begin{equation}
p(H)=p(H|HH)p(HH)+p(H|TT)p(TT)+p(H|HT)p(HT)+p(H|TH)p(TH)
\end{equation}
It is uncontroversial that $p(H|HH)=1$ and $p(H|TT)=0$. Given that one of the
outcomes is H and another T, Beauty knows that there are three awakenings: one
H and two T. Therefore, the conditional credences of Beauty for these cases
are $p(H|HT)=p(H|TH)={\frac{1}{3}}$. According to Lewis, Beauty has equal
credence for all possible outcomes of the coin tosses:
$p(HH)=p(TT)=p(HT)=p(TH)={\frac{1}{4}}$. It then follows from (1) that
Beauty's credence for H on awakening during the two weeks is $p(H)={\frac
{5}{12}}$. This is in contradiction with the
assumption that there should be no change between Elga's one-week
experiment and the similar two-week experiment. Therefore, unless Lewis rejects this very natural assumption, his approach is inconsistent.

On the analysis that I favour there is no such difficulty. On awakening,
Beauty's credences for the four outcomes of the coins tosses should not be
identical, they should be weighted according to the number of awakening
corresponding to these outcomes. Thus $p(HH)={\frac{1}{6}}$, $p(TT)={\frac
{1}{3}}$, and $p(HT)=p(TH)={\frac{1}{4}}$. Using (1), it follows that
$p(H)={\frac{1}{3}}$, just as in the one-week experiment in accordance
with Elga's argument.

\subsection{The quantum coin experiment}

An entirely independent argument to the same conclusion follows from the
interpretation of probability in the Many-Worlds Interpretation (MWI) of
quantum mechanics (an elaboration of the Everett approach (1957)). Consider
the toss of a quantum coin, say, for \ a simple example, the observation of a
photon, incident on semi-transparent mirror, as either reflected (R) or
transmitted (T). According to the MWI, the world splits in two: one (the
R-world) in which the photon is observed as reflected, and the other (the T
world) in which the photon is observed as transmitted.

Elga's experiment, but with such a quantum coin, is very similar to the
``sleeping pill'' experiment (Vaidman 1998), which was introduced to
give a possibility of an
ignorance interpretation of probability in the framework of the MWI.
 According to this approach, the observer assigns
the probability for outcomes of a quantum measurement in proportion to
the ``measures of
existence'' (Vaidman, 1998) of the corresponding worlds, the modulus squares of the
amplitudes of the corresponding branches of the universal wave
function. (See Saunders (1998) for a discussion of a similar concept
of ``measure'' .) There
is no direct meaning for this probability for the person who is going to
perform the quantum measurement (and who is put  to sleep for its duration):
there is no information, centered or uncentered, that he is then ignorant of.
The meaning for probability is given through the (identical) credences of the
two successors of the experimenter, on awakening, in centered propositions,
namely the propositions ``I am in the R-world'' and ``I am in the T-world''.
For each of these successors, there is a fact of the matter as to the outcome
of the experiment and he is ignorant of this fact.

 It is  worth remarking that the  problem  of the MWI recently posed by Peter
 Lewis does not arise in this approach. He argued (Lewis 2000) that 
 a believer in the MWI should  agree to play ``quantum Russian
roulette'', provided that death is instantaneous . The large
``measures'' of the worlds with dead successors is a good reason not to play.

There is a difference between Elga's experiment and the ``sleeping pill''   quantum coin flipping.
On awakening, Beauty is not only ignorant of which world she is in,
she is also ignorant of which time she is at in T-world. There are
three mutually exclusive propositions: ``I am in an H-world on a
Monday'', ``I am in a T-world on a Monday'' and ``I am in a T-world on
a Tuesday'', and she must assign credences in them summing to unity.
Failing any other information which could discriminate between the
three cases, her credences should be in proportion to the ``measures''
of the corresponding worlds, which happened to be exactly the same.
Therefore, her credence in each should be the same, namely one third.

There is an important difference between Elga's experiment with a
quantum and a classical coin . Classically, Beauty's credence concerns
a proposition, ``the coin is H'', and this along with the entire
sequence of events are located in a unique world. It is for this
reason that the proposition is reckoned to be uncentered. But using a
quantum coin, on the MWI, that is no longer so.  Quantum mechanical
outcomes are  in different worlds, and propositions
about such outcomes, such as ``the photon is T'' should be read as
tacitly indexical, ``in \textit{this} world the photon is T''; they
are properly speaking centered propositions.  To revert to our
previous discussion, the difference between case (i) and case (ii),
using a quantum coin, does not concern the centeredness or otherwise
of the proposition Beauty's credence is about. In both cases it is
centered.

On switching to a quantum coin in Elga's experiment, and on
interpreting quantum mechanics in terms of the MWI, one loses the
unusual feature that Lewis found so objectionable: that Beauty must
change her credence in an uncentered proposition, although her
uncentered evidence has not changed. Lewis may even \textit{agree}, in
this case, that Beauty's credence should change, for it is credence in
a centered proposition - and if so he will presumably agree that her
credence in H is one third.  But the quantum coin is considered to be
the best possible implementation of a fair coin: Beauty should not
have different credences in classical and quantum experiments.

\section
 {Byrne and Hall on Everett and Chalmers}

\subsection{Introduction}

   Byrne and Hall (1999) criticized the argument of Chalmers (1996) in
  favor of the Everett-style interpretation. They claimed to show ``the
  deep and underappreciated flaw in {\em any} Everett-style
  interpretation''. I will argue that it is possible to interpret
  Chalmers's writing in such a way that most of the criticism by Byrne
  and Hall does not  apply. (Recently I have learned that Chalmers
  himself (2000) partly accept the criticism, so my interpretation of
  his writing might differ from his original proposal.) In any case the
  general criticism of Byrne and Hall
  of the many-worlds interpretation is unfounded. The recent recognition that
  the Everett-style interpretations are good (if not the best)
  interpretations of quantum mechanics has, therefore, not been
  negated.

 It is probably impossible to present an interpretation of  quantum
 mechanics in unambiguous way  without writing equations.  Chalmers's presentation of
 Everett-style interpretation also can be understood in different
 ways. Instead of equations Chalmers used some technical jargon of
 quantum theory, however, some words like ``substates'' have no clear
 meaning even for physicists. Byrne and Hall (BH) interpreted
 Chalmers's jargon in a way which leads to contradictions. In this
 note I will argue that by taking a more positive approach, one can see
 in Chalmers's writing a consistent (although not necessarily very
 persuasive) argument.
 
 In the second part of their paper BH claimed to show not only that
 Chalmers has failed to establish his Everett-inspired interpretation,
 but that ``anything resembling it should not be taken seriously''.
 Their first point is of a general character: if the spaces of states
 in two theories are identical but the dynamics is not, it is not
 obvious that the interpretation of these states in the two theories
 must be identical too. BH point out that this is the situation
 regarding the interpretation of quantum states in the orthodox and
 the Everett interpretations. I will argue that although their general
 argument is correct, its application is not.  There is enough
 similarity between the dynamics that makes the identification
 plausible. The second point of BH is that the Everett-style
 interpretation has less ``substantive content'' than the orthodox
 interpretation. This is because in the Everett (many-worlds)
 interpretation there is no counterpart of ``outcome probabilities'',
 the concept of the orthodox interpretation associated with a system
 in a superposition of eigenstates of some variable.  I will argue
 that the  definition of the probability of an
 outcome in the framework of the many-worlds interpretation which I
 recently proposed solves this difficulty and makes this BH criticism obsolete.

  In Section 3.2 I will
 adopt the BH interpretation of Chalmers and will show (in a different
 from BH way) how it leads to a contradiction. In Section 3.3 I propose
 an alternative interpretation of Chalmers's writing which leads to a
 consistent argument. In Section 3.4 I critically analyze the general
 arguments of BH against the Everett-style interpretations. Finally,
 in Section 3.5 I summarize my defense of the many-worlds
 interpretation.

\subsection{Byrne and Hall interpretation and a contradiction in the Chalmers argument}

 The central thesis of Chalmers quoted by BH is the principle of
 {\it organizational preservation under superposition}:
 \begin{quotation}
\noindent
OPUS\break
   If a computation is implemented by a system in a maximal
   physical state $P$, it is also implemented by a system in a
   superposition of $P$ with orthogonal physical states.(Chalmers, 350)
 \end{quotation}
 Consider a simple model: a computer which performs calculations in a
 classical way. If at time $t_0$ the computer receives a classical
 input (a particular punching of its keyboard), then it evolves in
 time is such a way that it is always in a ``classical'' state. This
 means that all the registers of the computer at all times are in some
 definite states (exited or not exited) i.e., not in a superposition
 of excited and not excited.  Suppose that $P$ corresponds to a
 computation of a square of a number 5, while $Q$ corresponds to a
 computation of a square of a number 10. Denote $|P(t)\rangle$ a
 quantum state of the computer at time $t$ performing the calculation
 of the square of 5, while $|Q(t)\rangle$ a quantum state of the
 computer at time $t$ performing the calculation of the square of 10.
 In the two computations at any time the registers must be in
 different states, therefore, $|P(t)\rangle$ is orthogonal to
 $|Q(t)\rangle$. Thus, according to OPUS the computer in a quantum
 state
 \begin{equation}
   \label{r+}
  |R_+(t)\rangle \equiv 1/{\sqrt 2}
 (|P(t)\rangle + |Q(t)\rangle), 
 \end{equation}
  also implements computation of the
 square of 5. The quantum state 
 \begin{equation}
   \label{r-}
  |R_-(t)\rangle \equiv 1/{\sqrt 2}
 (|P(t)\rangle - |Q(t)\rangle), 
 \end{equation}
 is orthogonal to $|R_+(t)\rangle$. BH read Chalmers in such a way
 that OPUS can be applied to $|R_+(t)\rangle$ and $|R_-(t)\rangle$,
 i.e., that the superposition $1/{\sqrt 2} (|R_+(t)\rangle -
 |R_-(t)\rangle)$ also implements computation of the square of 5. But,
 \begin{equation}
 {1\over{\sqrt 2}} (|R_+(t)\rangle - |R_-(t)\rangle)=
{1\over 2}[ (|P(t)\rangle + |Q(t)\rangle) - (|P(t)\rangle -
|Q(t)\rangle)] = |Q(t)\rangle .
\end{equation} 
The state  $|Q(t)\rangle$ corresponds to the computation of the square of
10. It corresponds to the punching of a different input, it has
different registers activated during the calculation, it has different
output. Clearly, it does not implement computation of the square of 5.

Applying this direct reading of Chalmers, BH reached somewhat
different contradiction which lead them to reject Chalmers's approach.

\subsection{An alternative interpretation of Chalmers}

It is possible to read Chalmers in another way such
that the contradictions of the type described in the previous section do not arise. Let us make the following
modification of the OPUS principle:
 \begin{quotation}\noindent
   OPUS$'$\break 
If a computation is implemented by a system in a
   maximal physical state $P$ {\it which is not a superposition}, it
   is also implemented by a system in a superposition of $P$ with
   orthogonal physical states. 
 \end{quotation}
This modified principle can be applied to $P$ and $Q$, but it cannot
be applied  to $R_+$ and $R_-$ and, therefore, one cannot reach the
contradiction described above as well as  the contradictions
described by BH.

One might see that OPUS$'$ is what Chalmers actually had in mind even
though he did not say it explicitly. Indeed, another way to see the
difference between OPUS (as read by BH) and OPUS$'$ is that in the
latter it is required that $P$ corresponds to a {\it single}
experience.
Chalmers's first {\it definition} of the OPUS principle is:
\begin{quotation}
  If the theory predicts that a system  in a maximal physical state $P$
gives rise to an associated maximal phenomenal state $E$, then the
theory predicts that a system in a superposition of $P$ with some
orthogonal physical states will also give rise to $E$. (349)
\end{quotation}
The word ``associated'' hints that Chalmers meant that there is only
one experience (``phenomenal state $E$'' in Chalmers's notation) corresponding to 
physical state $P$.

In fact, BH saw a possibility of reading OPUS as OPUS$'$. The ``(Version
of) OPUS'' described in their section 5.2.3 is essentially OPUS$'$. They
rejected this because they understood that Chalmers denies the
existence of 
 {\it preferred basis}. BH are
correct in their criticism that without preferred basis there is no
way to distinguish between quantum state which is a ``superposition''
and a state which is not a ``superposition''.  Thus, the modification
of OPUS to OPUS$'$ cannot be done without assuming preferred basis. 

We can read Chalmers in such a way that we do not run into
inconsistency: Chalmers only objects to the claim that the
{\it mathematical} formalism of quantum mechanics, i.e. the Schr\"odinger
equation, leads to preferred basis. He cannot object to the existence
of preferred basis, but he  views it as arising from his theory of
consciousness. This reading of Chalmers is justified by the following
quotations:
\begin{quotation}
  Everett assumes that a superposed brain state will have a number of
  distinct subjects of experience associated with it, but he does
  nothing to justify this assumption. It is clear that this matter
  depends crucially on a theory of consciousness. A similar suggestion
  is made by Penrose (1989): ``... a theory of consciousness would be
  needed before the many-worlds view can be squared with what one
  actually observes'' (348)
\end{quotation}

\begin{quotation}
  ... last three strategies are all {\it indirect} strategies,
  attempting to explain the discreteness of experience by  explaining
  an underlying discreteness of macroscopic reality. An alternative
  strategy is to answer the question about experience {\it directly}. (349)
\end{quotation}

Before discussing quantum mechanics, Chalmers argues for a {\it
  principle of organizational invariance}:
\begin{quotation}
\noindent
POI\break
  Given any system that has conscious experiences, then any system
  that has the same fine-grained functional organization will have
  qualitatively identical experiences. (249)
\end{quotation}

The main difficulty which  BH see in  putting together the {\it
  principle of organization invariance} together with OPUS follows
from the same misinterpretation of Chalmers. If there is no  preferred basis
then they have reasons to say:
\begin{quotation}
 ... perceptual experience is (more or less) {\em entirely illusory}. 
 When you seem to see a voltmeter needle pointing to `10' your
 perceptual experience is probably veridical: the needle (if, indeed,
 we can sensibly speak of such a thing) is not pointing to `10' or
 anywhere else.
\end{quotation}
However, accepting preferred basis, even if it is  defined by the concept of
experience itself, resolves the difficulty: the pointer does point to
`10' and in addition, in parallel worlds, to other values too.

Chalmers claims that his {\em independently motivated} theory of
consciousness {\it predicts} that even in the world which is in a
giant superposition there are subjects who experience a discrete
world. He bases his argument on ``the claim that consciousness arises
from implementation of an appropriate computation.'' Taking the model
of a simple computer presented above, we can follow (at least
approximately) his proof on p. 350.  Projection of the superposed
state on ``the hyperplane of $P$'' might mean projection of the
quantum state of the computer in a ``superposed'' state at the initial
time on the state corresponding to the input of calculating square of the
number 5 which leads to quantum states of the various registers at
later times corresponding to this calculation.  The parallel between
the calculation and experience yields the desired result, but
accepting this parallel is relying  on our experience. So, if
we read Chalmers as BH do, that he claims to {\it deduce} ``what the world
is like if the Schr\"odinger equation is all'' without the guide of our
experience, then they have a valid criticism.  However, Chalmers
admits that Schr\"odinger equation cannot be all:
\begin{quotation}
... the only physical principle needed in quantum mechanics is the
Schr\"odinger equation, and the measurement postulate and other basic
principles are unnecessary baggage. To be sure, we need psychophysical
principles as well, but we need those principles in any case, and it
turns out that the principles that are plausible on independent
grounds can do the requisite work here. (350-351)
\end{quotation}
I feel that these ``independent grounds'' are  connected with our
experience in a stronger way than one might imagine  reading Chalmers. But
this fact cannot lead to rejection of this approach as BH  claim.

\subsection{ Byrne and Hall against {\it any} Everett-style interpretation}

BH start their argument by pointing out  that
the orthodox quantum theory and the Everett interpretation formally
defined on the same ``family of state spaces'' and that the difference
is  only
in dynamics.
Then they say  that because of the difference in dynamics it 
does not follow that the  quantum state corresponding to a particular
experience in the orthodox theory  will correspond to the same belief
(if at any) in the framework of the Everett theory. 

This might be considered as a criticism of  Chalmers if one reads
him as saying that Everett theory {\it predicts} what our experiences
should be, but usually this connection is {\it postulated} in
Everett-style theories. There is a strong motivation for this
postulate. The orthodox theory is defined only on a (tiny) part of the
space of all quantum states: macroscopic quantum systems cannot be in
a ``superposition states''. The dynamics of the allowed states between
quantum measurements is {\it identical} to the dynamics of the quantum
states in the Everett theory.
Let us discuss the example analyzed by BH at the end of p.385.  When a state
$\phi$ is a state of an observer who has the belief that the
measurement outcome was ``up'' in the orthodox theory, the dynamics
will tell that she will write  ``up'' in her lab-book. The
dynamics of the state $\phi$ in the Everett theory leads to the same
action. This justifies considering $\phi$ to be a ``belief vector'' in
the Everett theory  too.

BH proceed with their criticism claiming that  Everett's
interpretation has less of ``substantiative content'' because when a
quantum system is in a superposition of eigenstates with different
eigenvalues of some quantity {\bf M}, the orthodox interpretation
associates probabilities to the various outcomes, while the Everett
theory does not.

It is true that there is a difficulty with the concept of probability
in the framework of the Everett-style interpretation. The Everett
theory is a deterministic theory and it does not have a genuine
randomness of the collapse of the orthodox interpretation. A
deterministic theory might have the concept of {\it ignorance}
probability, but it is not easy to find somebody who is ignorant of
the result of a quantum experiment: it is senseless to ask what is the
probability that an observer will obtain a particular result, because
she will obtain {\it all} results for which there are a non-zero
probabilities according to the orthodox approach. It seems also
senseless to ask what is the probability of the observers in various
branches (these are persons with the same name and the same memories
about events which took place before the measurements, but who live in
different branches corresponding to the different outcomes) to obtain
various results, since obviously the probability to obtain the result
``${\rm \bf M} = m_i$'' in branch ``$j$'' is 1 if $i=j$ and it is 0 if
$i\neq j$. These are not the quantum probabilities we are looking for.

However, BH cannot dismiss the Everett-style
interpretations without even discussing current proposals to deal with this
problem (Lockwood {\it et al.} 1996, Saunders 1998, Deutsch 1999, etc.) Here, I
will sketch my proposal for solving this difficulty (Vaidman 1998,
2000). The splitting into various branches occurs usually before the
time when the observers in these branches become aware of the outcome
of the measurement. (To ensure this we may ask the observer to keep
her eyes close during the measurement.) Thus, an observer in each
branch is ignorant about the outcome of the measurement and she can
(while any external person cannot!) define the the {\it ignorance}
probability for the outcome of the measurement.  She will do so using
standard probability postulate: the probability of an outcome is
proportional to the square of the amplitude of the corresponding
branch. Moreover, since observers in {\it all} these branches have
identical concept of ignorance probability and since they all are
descendants of the observer who performed the experiment, we can
associate probability for an outcome of a measurement for this
observer in the sense that this is the common ignorance probability of
her descendants in various branches.

The fact that I have used a probability postulate here does not spoil
the argument: I had to show that substantive content of Everett
interpretation is not less than that of the orthodox interpretation.
The latter has the probability postulate as well. What was done here
(and what was not trivial from the beginning) was to present a way
which allows to {\it define} probability in the frame of the
many-worlds interpretation. This definition also resolves the
difficulty recently discussed by Lewis (2000). He argued that a
believer in the many-worlds (minds) interpretation should agree to
play a ``quantum Russian Roulette'' provided the death is
instantaneous. Indeed, the instantaneity makes it difficult to
establish the probability postulate, but after it has been justified
in the wide range of other situations it is natural to apply the
postulate for all cases.

The last argument of BH relies on their claim that Everett-style
interpretation lacks ``statistical algorithm''.  Since the ignorance probability defined above generates the same
statistical algorithm as the the orthodox theory, this argument does not
hold either.

\subsection{Summary}

The main claim of BH is ``that {\it any}
 Everett-style interpretation should be rejected''. The basis of their
 argument is the observation that neither Chalmers nor anybody else
 can answer the question: ``What the world is like if the Shr\"odinger
 equation is all?'' It is true that this question is much more
 difficult to answer in the framework of the Everett-style
 interpretation relative to interpretations which do not have
 multitude of worlds. ``The world is everything which exist'' is not a
 valid definition. Moreover, the Shr\"odinger equation itself cannot
 define the concept of a ``world''.  The world is the concept defined
 by conscious beings and it requires the analysis of the mind-body
 connection.  Chalmers's theory of consciousness provides an answer.
 One might argue how substantial his answer is, but even if there is
 no a detailed answer to this question today, one cannot  reject the
 Everett interpretation. It suffices  that 
 Everett's theory is consistent with what we see as our world. It is so
 superior to the alternatives from the physics point of view, because
 it avoids randomness and action at a distance in Nature (e.g., see
 Vaidman 2000), that it is still preferable in spite of the fact that
 it is less satisfactory from the philosophical point of view.
  Therefore, even if BH were able to point out a 
  difficulty in obtaining the interpretation out of the ``bare theory''
 this would not be enough for rejecting the Everett interpretation.
Moreover, I have argued that the  BH have not presented persuasive
arguments  showing  the difficulty. Their first
 argument is that it is not obvious that the correspondence between
 quantum states and classical properties in the orthodox quantum
 mechanics can be transformed as it is to the Everett
 interpretation. This argument does not take into account 
 the similarity in dynamics which justifies the identification.
 Their other arguments rely on the well known difficulty in the
 interpretation of probability in the many-worlds interpretation 
 disregard recently proposed solutions of this problem.

In summary, BH were not able to show a flaw in Everett-style
 interpretations.  The temptation to appeal to the philosophy of mind
 in interpreting quantum mechanics, in particular, the idea that a
 theory of mind might help rescue from the difficulties with standard
 interpretation is still very attractive. Indeed, the Everett-style
 interpretation which says that physics is described in full by the
 Schr\"odinger equation is the most satisfactory  from the physics point
 of view. What is left is to complete Chalmers's work, i.e. to
 elaborate the connection between the quantum state evolving according to
 the Schr\"odinger equation and our experience.

\section*{Acknowledgments}It is a pleasure to thank David Chalmers,
Adam Elga, Christopher Fuchs, Andrei Khrennikov, Simon Saunders and
Ruediger Schackfor for illuminating comments. This research was
supported in part by grant 62/01 of the Israel Science Foundation and
the EPSRC grant GR/N33058.

\section*{References}

\vskip.13cm \noindent
 Aumann, R. J., Hart, S., and Perry, M. 1997 The
forgetful passenger, \textit{Games and Economic Behaviour} {\bf 20}: 117-120.

\vskip .13cm \noindent 
Byrne, A. and Hall, N. 1999. Chalmers on Consciousness and Quantum
Mechanics, {\it Philosophy of Science} {\bf 66}: 370-390.

\vskip .13cm \noindent 
     Caves, C. M.,   Fuchs, C. A., and    Schack, R. 2001
 Making good sense of quantum probabilities, e-print: http://xxx.lanl.gov/quant-ph/0106133.

\vskip .13cm \noindent 
Chalmers, D. J. 1996
{\it The Conscious Mind},  New York: Oxford University Press. 

\vskip .13cm \noindent 
Chalmers, D. J. 2000
  Responses to articles on my work, {\it www homepage}, http://www.u.arizona.edu/~chalmers/responses.html

\vskip .13cm \noindent 
Deutsch D. 1999 Quantum Theory of Probability and Decisions, {\it  Proceedings of the Royal Society of London}  {\bf A 455}:
    3129-3137.

\vskip.13cm \noindent
 Elga, A. 2000 Self-locating belief and the Sleeping
Beauty problem \textit{Analysis} {\bf 60}:143-147.

\vskip.13cm \noindent
 Everett, H. 1957 ``Relative State' Formulation of
Quantum Mechanics,'' \textit{Review of Modern Physics} {\bf 29}: 454-462.

\vskip.13cm \noindent
 Lewis, D. 2001 Sleeping Beauty: reply to Elga,
\textit{Analysis} {\bf 61}: 171-176. 

\vskip.13cm \noindent
 Lewis, P. J. 2000  What is it like to be
Schr\"{o}dinger's cat? \textit{Analysis} \textbf{60}: 22-29.

\vskip .13cm \noindent 
 Lockwood, M., Brown, H. R., Butterfield, J., Deutsch, D., Loewer, B., Papineau, D., Saunders, S. 1996. Symposium: The
    `Many Minds' Interpretation of Quantum Theory, {\it British Journal for the Philosophy of Science}  {\bf 47}: 159-248. 

\vskip.13cm \noindent
 Penrose, R., 1989
{\it The emperor's new mind : concerning computers, minds, and the laws of physics},
           New York: Oxford University Press.

\vskip.13cm \noindent
 Saunders, S. 1998
 Time, Quantum Mechanics, and
Probability, \textit{Synthese} \textbf{114}: 373-404.

\vskip.13cm \noindent
 Sklar, L. 1993
 { \textit{Physics and Chance :
Philosophical Issues in the Foundations of Statistical Mechanics}, Cambridge :
Cambridge University Press, 1993. }

\vskip.13 cm\noindent
 Vaidman, L. 1998 On Schizophrenic Experiences of the
Neutron or Why We should Believe in the Many-Worlds Interpretation of Quantum
Theory, \textit{International Studies in the Philosophy of Science}
\textbf{12}: 245-261.

\vskip .13 cm\noindent
 Vaidman, L. 2001a  Tests of Bell Inequalities,
 L. Vaidman 
{\it Physics Letters} {\bf A 286}: 241-244.

\vskip .13 cm\noindent
 Vaidman, L. 2001b The Many-Worlds
Interpretation of Quantum Theory, {\it Stanford Encyclopedia of
  Philosophy}, http://plato.stanford.edu/entries/qm-manyworlds/

\vskip .13 cm\noindent
 Vaidman, L. and Saunders, S. 2001 
On Sleeping Beauty Controversy, {\it  PhilSci Archive},
http://philsci-archive.pitt.edu/documents/disk0/00/00/03/24/index.html

\end{document}